\documentclass[aip,graphicx]{revtex4-1}
\usepackage{graphicx}
\usepackage{gensymb}
\begin{document}


\title{Evolution of the linear-polarization-angle-dependence of the radiation-induced magnetoresistance-oscillations with microwave power }


\author{Tianyu Ye}
\affiliation{Department of Physics and Astronomy, Georgia State University, Atlanta, Georgia 30303, USA}
\author{W. Wegscheider}
\affiliation{Laboratorium f\"ur Festk\"orperphysik, ETH Z\"urich, 8093 Z\"urich, Switzerland}
\author{R. G. Mani}
\affiliation{Department of Physics and Astronomy, Georgia State University, Atlanta, Georgia 30303, USA}

\date{\today}

\begin{abstract}
We examine the role of the microwave power in the linear polarization angle dependence of the microwave radiation induced magnetoresistance oscillations observed in the high mobility GaAs/AlGaAs  two dimensional electron system. Diagonal resistance $R_{xx}$ was measured at fixed magnetic fields corresponding to the photo-excited oscillatory extrema of $R_{xx}$ as a function of both the microwave power, $P$, and the linear polarization angle, $\theta$. Color contour plots of such measurements demonstrate the  evolution of the $R_{xx}$ versus $\theta$ line shape with increasing microwave power. We report that the non-linear power dependence of the amplitude of the radiation-induced magnetoresistance oscillations distorts the cosine-square relation between $R_{xx}$ and $\theta$ at high power.
\end{abstract}


\maketitle

\section{introduction}
The high mobility 2D electron system (2DES) exhibits zero-resistance states at the minima of  giant magnetoresistance oscillations induced by microwave and terahertz excitation at low magnetic fields and low temperatures.\cite{Maninature2002,ZudovPRLDissipationless2003} These radiation-induced zero resistance states differ from the well-known quantized Hall effect zero-resistance states by the necessity of photo-excitation for the observability of the effect and the absence of concurrent plateaus in the Hall effect. The huge photo-induced modulation of the dark magnetoresistance in this special photo-response of the 2DES yields potential applications in microwave and terahertz wave sensing.\cite{ManiPRBterahertz2013}

The amplitude of the radiation-induced magnetoresistance oscillations (RiMOs) depends on factors such as radiation frequency\cite{ManiPRLPhaseshift2004, ManiAPL2008, ManiPRBPhaseStudy2009}; temperature\cite{ManiEP2DS152004}, radiation power\cite{ManiPRBAmplitude2010}, linear polarization direction\cite{ManiPRBPolarization2011, RamanayakaPolarization2012}, angle between magnetic field and sample normal\cite{ManiPRBTilteB2005} and current through sample\cite{ManiPRBVI2004}. All these factors have been examined both experimentally\cite{KovalevSolidSCommNod2004,SimovicPRBDensity2005,SmetPRLCircularPolar2005,WiedmannPRBInterference2008,DennisKonoPRLConductanceOsc2009,ArunaPRBeHeating2011,ManinatureComm2012,TYe2013,Mani2013sizematter,ManiNegRes2013,TYe2014combine} and theoretically\cite{DurstPRLDisplacement2003,AndreevPRLZeroDC2003,RyzhiiJPCMNonlinear2003,KoulakovPRBNonpara2003,RyzhiiPRBConduct2003,LeiPRLBalanceF2003,DmitrievPRBMIMO2005,LeiPRBAbsorption+heating2005,InarreaPRLeJump2005,ChepelianskiiEPJB2007,Inarrea2008,FinklerHalperinPRB2009,ChepelianskiiPRBedgetrans2009,InarreaPRBPower2010,Inarrea2011,Inarrea2012,Lei2012Polar,Inarrea2013Polar,Kunold2013,Zhirov2013,Lei2014Bicromatic}. However, there are still open questions, two of which are the microwave power dependence and microwave polarization dependence of RiMOs. For the power dependence, there has been a debate whether the amplitude of RiMOs increases linearly with the microwave power as indicated by the inelastic model\cite{DmitrievPRBMIMO2005}. Here, some experimental work\cite{ManiPRBAmplitude2010,TYe2013,TYe2014combine} and the radiation driven electron orbital model\cite{InarreaPRBPower2010} have suggested  that the RiMOs’ amplitude actually increases sublinearly in the microwave power. So far as the role of microwave polarization in RiMOs is concerned, the inelastic model\cite{DmitrievPRBMIMO2005} and experiment\cite{SmetPRLCircularPolar2005} indicated that RiMOs are independent of the polarization angle of linearly polarized microwave, while another set of experiments\cite{ManiPRBPolarization2011, RamanayakaPolarization2012} and theory, i.e, the displacement model\cite{Lei2012Polar} and the radiation driven electron orbital model\cite{Inarrea2013Polar}, suggested that the RiMO amplitude depends on the polarization angle of linearly polarized microwave. Indeed,  ref. \cite{RamanayakaPolarization2012} has demonstrated a sinusoidal relation, i.e.,  $R_{xx}(\theta)=A \pm Ccos^2(\theta-\theta_0)$, between RiMO amplitude $R_{xx}$ and linear polarization angle $\theta$. Moreover, the results indicated that, at high radiation intensities, there was a systematic deviation in the polarization angle dependence from this cosine-square rule.\cite{RamanayakaPolarization2012}   This result, see Fig. 4 of ref. \cite{RamanayakaPolarization2012}, indicated RiMOs' polarization angle dependence is affected by the radiation intensity. The motivation for the present study has, therefore, been to understand the role of the microwave power/intensity in the linear-polarization-angle dependence of RiMOs.

\section{experiments and results}
\begin{figure}[t]
\centering
\includegraphics[width= 85mm]{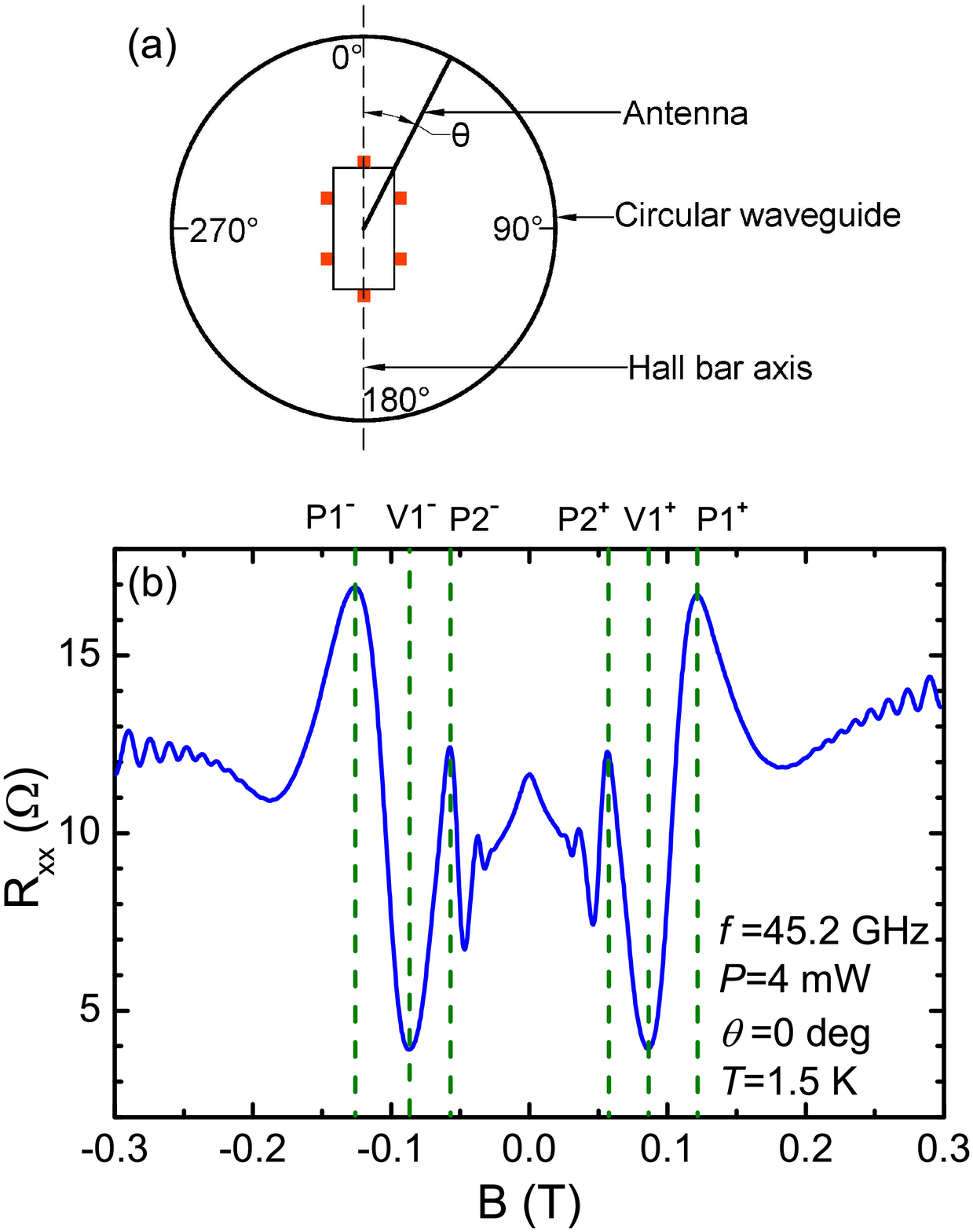}
\caption{(Color online) (a)A sketch of the polarization orientation in the magnetotransport measurement. Here, the antenna and the microwave launcher rotate clockwise with respect to the long axis of Hall bar sample to set the polarization angle $\theta$. (b)Diagonal resistance $R_{xx}$ versus the magnetic field $B$ with microwave photo-excitation at 45.2 GHz and $T=1.5$ K. The polarization angle, $\theta$, is zero. Symbols, i.e., P1$^-$, V1$^-$, etc., at the top abscissa  mark the magnetic fields of some of the peaks and valleys of the oscillatory magnetoresistance.}
\end{figure}

Experiments were carried out on  Hall bars fabricated from high mobility GaAs/AlGaAs heterojunctions. A long cylindrical waveguide sample holder with sample mounted at the end was inserted into a variable temperature insert (VTI), inside the bore of a superconducting solenoid. A temperature of 1.5K was realized by pumping on the liquid helium within the VTI insert. The specimens reached the high mobility condition after brief illumination with a red light-emitting-diode at low temperature. A commercially available microwave synthesizer was used to provide microwave illumination via a launcher at the top of sample holder. The linearly polarization angle is defined as the angle between the long axis of the Hall bar and the microwave antenna in the microwave launcher. This angle was changed by rotating the microwave launcher outside the cryostat, see Fig. 1 (a). Finally, standard low frequency $ac$ lock-in techniques were adopted to measure the magnetoresistance, $R_{xx}$. 

Figure 1 (b) exhibits a magnetic field sweep of $R_{xx}$ with photoexcitation at $f=45.2$ GHz and $P=4$ mW, at the linear  polarization angle $\theta= 0 \degree$ . The magnetoresistance curve exhibit RiMOs up to the third order on each side of the magnetic field axis. Here, the peaks and valleys are denoted by ``P'' and ``V''. Thus, for instance, P1 and V1 indicate the first prominent peak and valley, respectively, from the high magnetic field direction. Further, ``+'' ( ``$-$'') indicate the positive (negative) magnetic field. In Figs. 2 - 4 , $R_{xx}$ will be reported at the fixed magnetic fields corresponding to P1$^-$, V1$^-$, P2$^-$, P1$^+$, V1$^+$ and P2$^+$, with power sweeps at discrete polarization angles in the range between $0 \degree$ and $360 \degree$ with a $10 \degree$ increment.

\begin{figure}[t]
\centering
\includegraphics[width=85 mm]{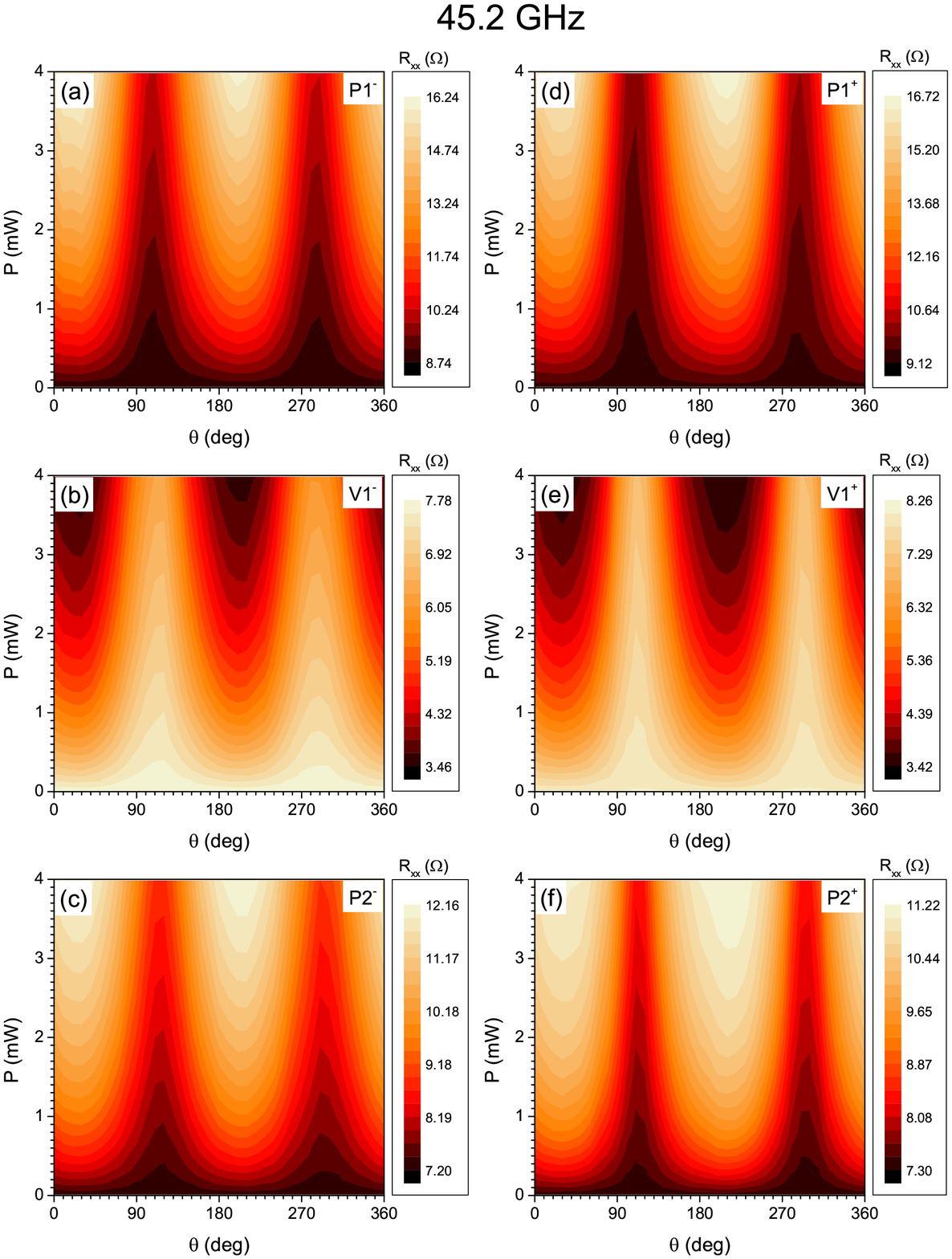}
\caption{(Color online) Color contour plots of diagonal resistance $R_{xx}$ as a function of both microwave power (ordinates) and polarization angle (abscissas) at $f= 45.2$ GHz and at the magnetic field corresponding to (a) P1$^-$, (b) V1$^-$, (c) P2$^-$, (d) P1$^+$, (e) V1$^+$ and (f) P2$^+$. $R_{xx}$ values are indicated by the color scales on the right side of each figure.}
\end{figure}

To demonstrate the evolution of microwave polarization dependence, color contour plots are exhibited in figure 2. Here, the abscissas represent the linear polarization angle, $\theta$, and the ordinates represent the microwave power, $P$. Color scales represent numerical values of diagonal resistance $R_{xx}$: a warmer color indicates a higher resistance and a darker color means lower resistance. Numerical color scales for each plot in figure 2 are shown adjacent to the plots. Fig. 2(a) to (f) are the plots for P1$^-$, V1$^-,$ P2$^-$, P1$^+$, V1$^+$ and P2$^+$, as indicated. For each plot, the phase shift angle $\theta_0$\cite{RamanayakaPolarization2012} is approximately at $30 \degree$, and there is no obvious dependence of the phase shift upon the extrema or magnetic field. From the color gradient at a fixed polarization angle, it is clear that the resistance values are monotonically increasing as microwave power increases at P1$^-$, P2$^-$, P2$^+$ and P1$^+$, and monotonically decreasing as microwave power increase at V1$^-$ and V1$^+$. At a constant $P$, resistance changes periodically with $\theta$.

\begin{figure}[t]
\centering
\includegraphics[width= 150mm]{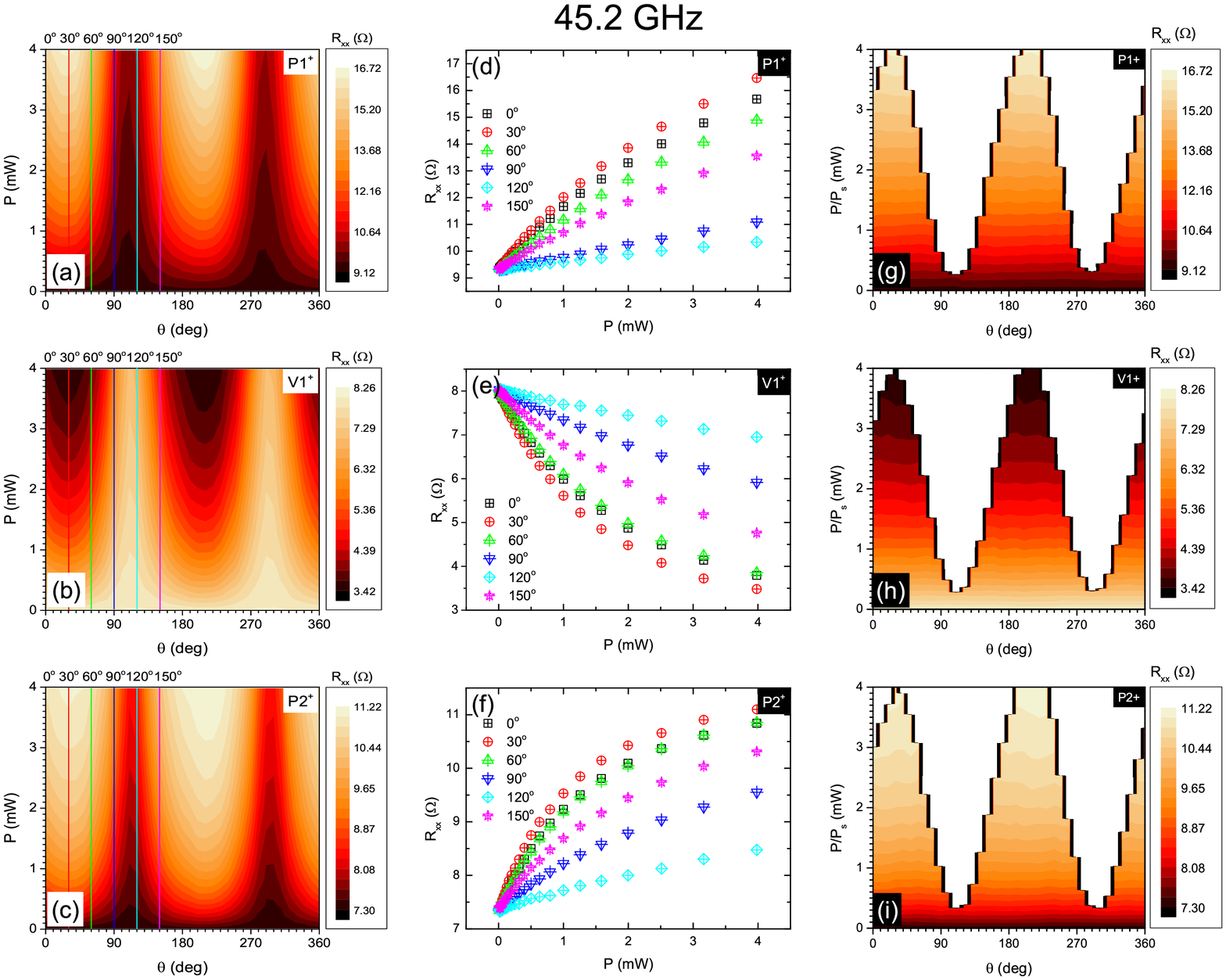}
\caption{(Color online) Figures in the left column are the color contour plots of diagonal resistance $R_{xx}$ as a function of microwave power and polarization angle at $f=45.2$ GHz and at the magnetic field corresponding to (a) P1$^+$, (b) V1$^+$ and (C) P2$^+$. Vertical solid lines in each figure indicate the polarization angles, at which $R_{xx}$ vs $P$ profile curves are showed in the middle column figures: (d) for P1$^+$, (e) for V1$^+$ and (f) for P2$^+$. The line-color in the left column should be matched to the same color symbols in the center column. Right column exhibits contour plots with normalized ordinate scales. Here, $P/P_s$ are used for ordinates in (g) for P1$^+$, (h) for V1$^+$ and (i) for P2$^+$.}
\end{figure}

Figure 3 shows the contour plots of P1$^+$, V1$^+$ and P2$^+$ on left column. The vertical lines in the color plots indicate certain selected discrete microwave polarization angles from $0 \degree$ to $150 \degree$. $R_{xx}$ is plotted vs. $P$ at these selected polarization angles in the middle column of Fig. 3. At these different polarization angles, the $R_{xx}$ vs $P$ curves initiate from the same point at 0 mW but start to separate with increasing $P$. For P1$^+$ and P2$^+$, $R_{xx}$ increases as the power increases, and $R_{xx}$ decreases as power increases at V1$^-$. In all cases, the change in $R_{xx}$ with $P$ is non- linear. Although $R_{xx}$ vs $P$ curves for different polarization angles separate once microwave power start to increase from zero, they could still be normalized to the same curve with dividing abscissa by a scaling factor $P_s$\cite{TYe2014combine}. Such scaling factors are different for different polarization angles. The reciprocal scaling factor is also a cosine square function of linear polarization angle, and it follows $1/P_s(\theta)=A+Ccos^2(\theta-\theta_0)$\cite{TYe2014combine}. In the right column of Fig. 3, the ordinate-scales constitute a normalized scale, which means the microwave power scale is divided by the scaling factors that belong to each polarization angle. Although experiments were carried out with $0 \le P \le 4$ mW, after normalization, for some polarization angles, the normalized power would cover a smaller range because the scaling factor $P_s$ can exceed unity. Since, here all the power scales are normalized with respect to the one for $\theta=30 \degree$, $P/P_s$ spans full range at $30 \degree$ and $210 \degree$. The normalization has the effect of compressing data in the ordinate direction away from $\theta = 30 \degree$ and $\theta=210 \degree$. As a result, white spaces appear in the color plots of the right column.  In Fig. 3 (g) to (i), the color contours exhibit a horizontal orientation, which indicates that after normalization, $R_{xx}$ takes on the same values for different polarization angles. Interestingly, the profile or envelope of $P/P_s$ vs $\theta$  in the contour plots of Fig. 3(g) to (i) are sinusoidal and the peaks appear at $30 \degree$ and $210 \degree$, which suggests that the profile curve can be described by $P/P_s(\theta)=A+Ccos^2(\theta - \theta_0)$.

\begin{figure}[t]
\centering
\includegraphics[width= 150mm]{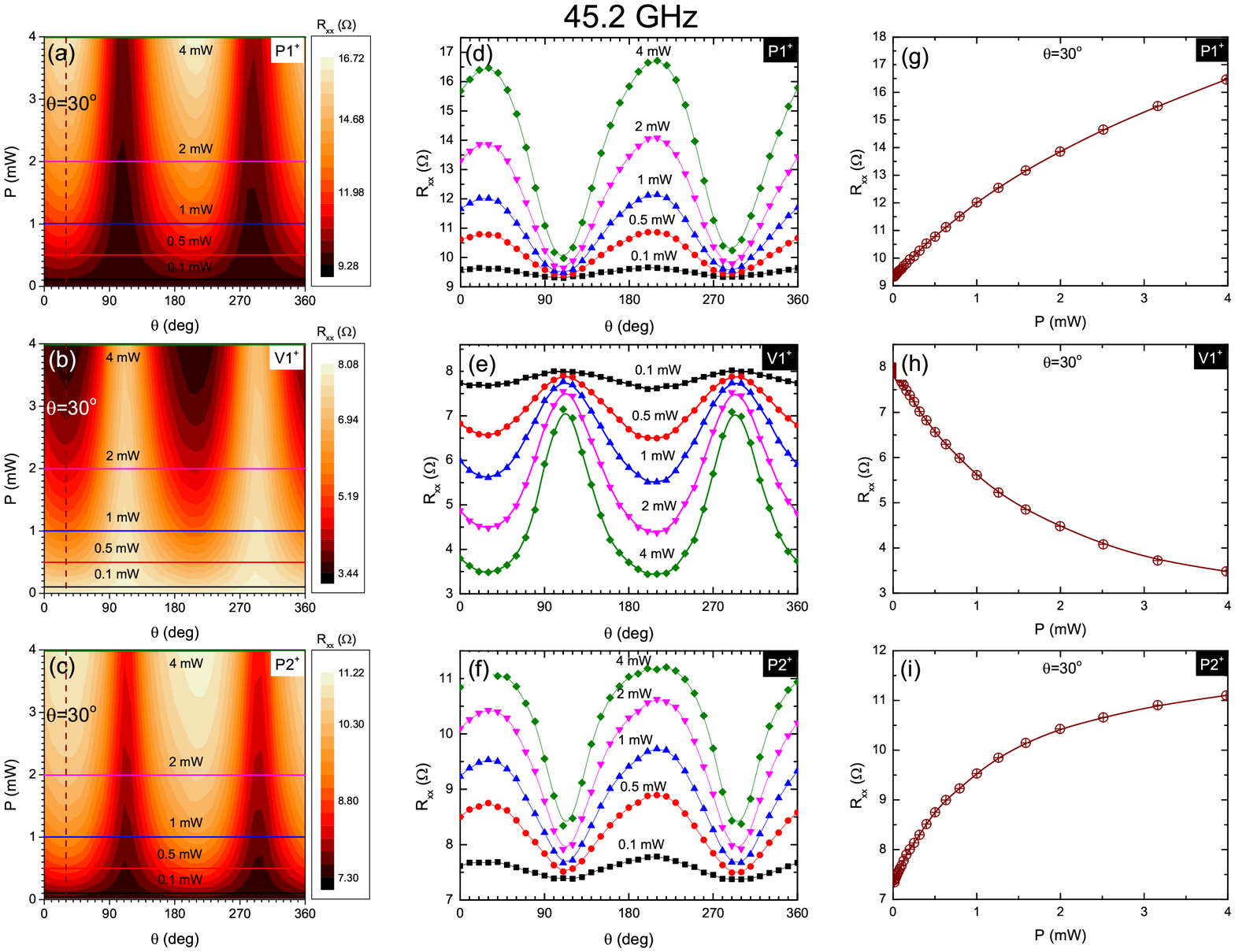}
\caption{(Color online) Figures in the left columns are the color contour plots of diagonal resistance $R_{xx}$ as a function of the microwave power and the polarization angle at $f=45.2$ GHz and at the magnetic field corresponding to (a) P1$^+$, (b) V1$^+$ and (C) P2$^+$. Horizontal lines in each figure indicate the  power levels, at which $R_{xx}$ vs $\theta$ profile curves are shown in the middle column figures: (d) for P1$^+$, (e) for V1$^+$ and (f) for P2$^+$. The line-color in the left column should be matched to the same color curves in the middle column. Vertical dashed lines in the left column indicate the phase shift angle $\theta_{0} = 30 \degree$. The $R_{xx}$ vs. $P$ at this angle are plotted in the right column in panels  (g), (h) and (i) for  P1$^+$, V1$^+$ and P2$^+$, respectively.}
\end{figure}

The  left column of Fig. 4 shows the same contour plots of P1$^+$, V1$^+$ and P2$^+$ as in Fig. 3. The horizontal solid lines in the color plots of Fig. 4 indicate discrete microwave powers from 0.1 mW to 4 mW. Extracted  $R_{xx}$ vs $\theta$ at these $P$ are shown in the middle column of Fig. 4. The plots in the middle column of Fig. 4 show a deviation from simple sinusoidal behavior with increasing $P$ as reported in \cite{RamanayakaPolarization2012}. Here, for P1$^+$ and P2$^+$ the valleys become sharper as microwave power increases from 0.1 mW to 4 mW and for V1$^+$ the peaks become sharper as the power increases. Simultaneously, with increasing $P$, the peaks of P1$^+$ and P2$^+$ become more broadened, and at 4 mW, the peak of P2$^+$ shows signs of forming a broad plateau-like structure. The valleys of V1$^+$ also become more broadened with increasing $P$. In comparing the deviation from sinusoidal behavior for P1$^+$ and P2$^+$, the  change in the $R_{xx}$ vs $\theta$ line shape is more obvious for P2$^{+}$ than P1$^+$, which suggests the $R_{xx}$ vs $\theta$ line shape evolution with microwave power becomes more pronounced at lower magnetic fields. So far as the phase is concerned,  in the left column contour plots,, vertical dashed lines point out phase angle, $\theta_0=30 \degree$. $R_{xx}$ vs. $P$ at this phase angle are plotted in the right panel of Fig. 4. In the middle column of Fig. 4, at $\theta=30 \degree$, $R_{xx}$ either increases (for P1$^+$ and P2$^+$) or decreases (for V1$^+$) as the microwave power increases, but the rate of change of $R_{xx}$ with $P$  are different, see also the $R_{xx}$ vs. $P$ plots in the right column of Fig. 4. It appears that  the different non-linearity at different polarization angles is the main cause for the distortion of $R_{xx}$ vs $\theta$ curves from simple sinusoidal form at high microwave powers.

\section{discussion}
We have examined the line shape evolution, or more specifically, the line shape distortion, of $R_{xx}$ vs $\theta$ observed with increasing $P$, especially at high microwave powers. The exhibited results indicate: i) microwave power and linear polarization angle both influence the oscillatory $R_{xx}$ in the high mobility 2DES in the regime of the radiation-induced magnetoresistance oscillations. ii) At the peaks or valleys of the RiMOs, $R_{xx}$ is not a linear function of microwave power. iii) $R_{xx}$ vs $P$ curves at different polarization angles exhibit different non-linearity. However, they can be normalized to the same curve using an empirical power scaling factor $P_s$. This also suggests, see Fig. 3 (g) to (i), that the amplitude of RiMOs may be sensitive to an effective microwave power, which is a cosine square function of polarization angle. iv) The difference in the non-linearity of $R_{xx}$'s power dependence at different polarization angles is the factor that makes $R_{xx}$ vs $\theta$ deviate from cosine square function at high microwave powers.

Regarding the microwave power dependence and linear polarization dependence of RiMOs, previous studies\cite{InarreaPRBPower2010,ManiPRBAmplitude2010,RamanayakaPolarization2012,Lei2012Polar,Inarrea2013Polar,TYe2014combine} have also drawn the conclusion that at the magnetic fields corresponding to the extrema of RiMOs, the $R_{xx}$ varies nonlinearly with $P$  and the extremal $R_{xx}$ follows $R_{xx}(\theta) = A \pm Ccos^{2}(\theta-\theta_{0})$ at low $P$.  Regarding the polarization dependent distortion from cosine square function at high microwave power range, one must consider power and polarization angle together. Among the available theoretical models, few have considered both power dependence and polarization dependent at once. One model that considers them both is the radiation driven electron orbital model\cite{InarreaPRLeJump2005}, which describes a periodic back-and-forth radiation-driven motion of the electron orbits and the conductivity modulation resulting from the average scattering jump. Simulations based on this model indicate a non-linear power dependence of RiMOs\cite{InarreaPRBPower2010}, i.e., $A=A_0P^{\alpha}$, where $A$ is the amplitude of RiMOs, $A_0$ and $\alpha \approx 1/2$ are constants and $P$ is microwave power. Moreover, the polarization dependence simulation\cite{Inarrea2013Polar} appear to suggest a distorted cosine square function. It seems plausible that if microwave power and polarization angle are considered together, this theory might model the power and polarization dependence of microwave-induced oscillatory magnetoresistance. The displacement model\cite{DurstPRLDisplacement2003,RyzhiiJPCMNonlinear2003,LeiPRLBalanceF2003}, which describes microwave photo-excited electrons scattered by impurities, and gives rise to an additional current density due to radiation, also considered the polarization dependence as sinusoidal\cite{Lei2012Polar}. However, this simulation has not yet included the power dependence. Also, the microwave intensities used in this simulation are quite different from the experimental intensities. Another displacement model simulation\cite{RyzhiiPRBConduct2003} has also studied the microwave power dependence of RiMOs. They stated that photo-conductivity $\sigma_{ph}$, at extrema of RiMOs, is a non-linear function of microwave power in the electron transition between adjacent Landau levels. This non-linearity is apparent especially at high microwave power range. However, this displacement model simulation indicated neither the polarization dependence nor its distortion due to high power.  One might expect, however, to see the polarization dependent curve distortion at relative high microwave power in simulations utilizing the displacement model. Most other theoretical models have not predicted polarization dependence of RiMOs, not to mention the $R_{xx}$ vs ${\theta}$  line shape distortion due to high microwave power. 

\section{conclusion}
In conclusion, we have examined the evolution of the lineshape of $R_{xx}$ vs $\theta$ in RiMOs with $P$ using color contour plots of $R_{xx}$ as a function of both microwave power $P$ and linear polarization angle $\theta$. The different non-linearity of $R_{xx}$ vs. $P$ traces at different polarization angles is found to be the main factor that influences line shape distortion of the sinusoidal $R_{xx}$ vs. $\theta$ relation.  As well, this work support the non-linear power dependent and generally sinusoidal polarization dependence of RiMOs at low $P$.

\section{acknowledgement}
Magnetotransport measurements at Georgia State University are
supported by the U.S. Department of Energy, Office of Basic Energy
Sciences, Material Sciences and Engineering Division under
DE-SC0001762. Additional support is provided by the ARO under
W911NF-07-01-015.

\pagebreak

\section*{Figure Captions}
Figure 1:(a)A sketch of the polarization orientation in the magnetotransport measurement. Here, the antenna and the microwave launcher rotate clockwise with respect to the long axis of Hall bar sample to set the polarization angle $\theta$. (b)Diagonal resistance $R_{xx}$ versus the magnetic field $B$ with microwave photo-excitation at 45.2 GHz and $T=1.5$ K. The polarization angle, $\theta$, is zero. Symbols, i.e., P1$^-$, V1$^-$, etc., at the top abscissa  mark the magnetic fields of some of the peaks and valleys of the oscillatory magnetoresistance. 

Figure 2: Color contour plots of diagonal resistance $R_{xx}$ as a function of both microwave power (ordinates) and polarization angle (abscissas) at $f= 45.2$ GHz and at the magnetic field corresponding to (a) P1$^-$, (b) V1$^-$, (c) P2$^-$, (d) P1$^+$, (e) V1$^+$ and (f) P2$^+$. $R_{xx}$ values are indicated by the color scales on the right side of each figure.

Figure 3: Figures in the left column are the color contour plots of diagonal resistance $R_{xx}$ as a function of microwave power and polarization angle at $f=45.2$ GHz and at the magnetic field corresponding to (a) P1$^+$, (b) V1$^+$ and (C) P2$^+$. Vertical solid lines in each figure indicate the polarization angles, at which $R_{xx}$ vs $P$ profile curves are showed in the middle column figures: (d) for P1$^+$, (e) for V1$^+$ and (f) for P2$^+$. The line-color in the left column should be matched to the same color symbols in the center column. Right column exhibits contour plots with normalized ordinate scales. Here, $P/P_s$ are used for ordinates in (g) for P1$^+$, (h) for V1$^+$ and (i) for P2$^+$.

Figure 4:  Figures in the left columns are the color contour plots of diagonal resistance $R_{xx}$ as a function of the microwave power and the polarization angle at $f=45.2$ GHz and at the magnetic field corresponding to (a) P1$^+$, (b) V1$^+$ and (C) P2$^+$. Horizontal lines in each figure indicate the  power levels, at which $R_{xx}$ vs $\theta$ profile curves are shown in the middle column figures: (d) for P1$^+$, (e) for V1$^+$ and (f) for P2$^+$. The line-color in the left column should be matched to the same color curves in the middle column. Vertical dashed lines in the left column indicate the phase shift angle $\theta_{0} = 30 \degree$. The $R_{xx}$ vs. $P$ at this angle are plotted in the right column in panels  (g), (h) and (i) for  P1$^+$, V1$^+$ and P2$^+$, respectively.

\pagebreak

\begin{figure}[t]
\centering
\includegraphics[width= 85mm]{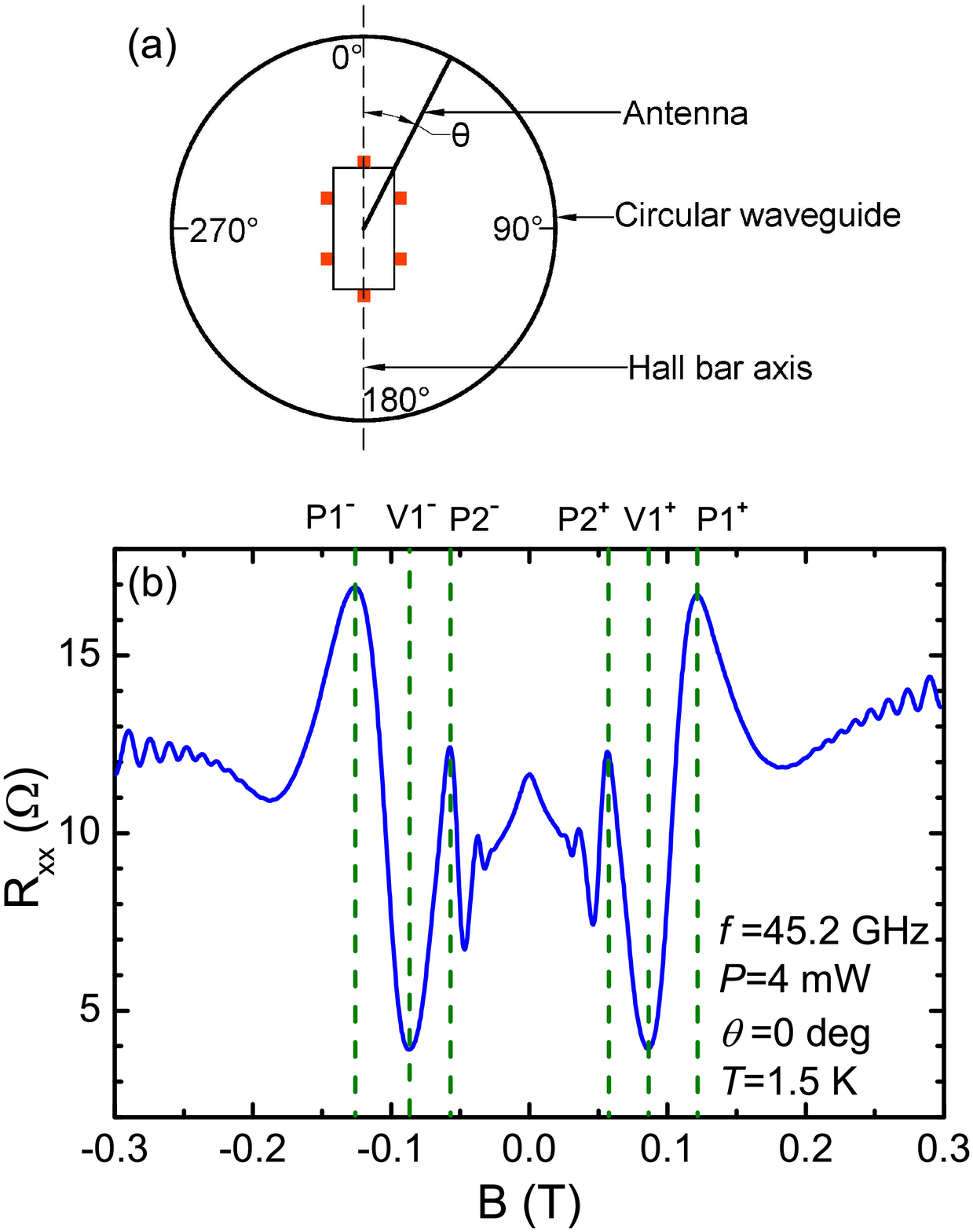}
\begin{center}
Figure 1
\end{center}
\end{figure}

\begin{figure}[t]
\centering
\includegraphics[width=85 mm]{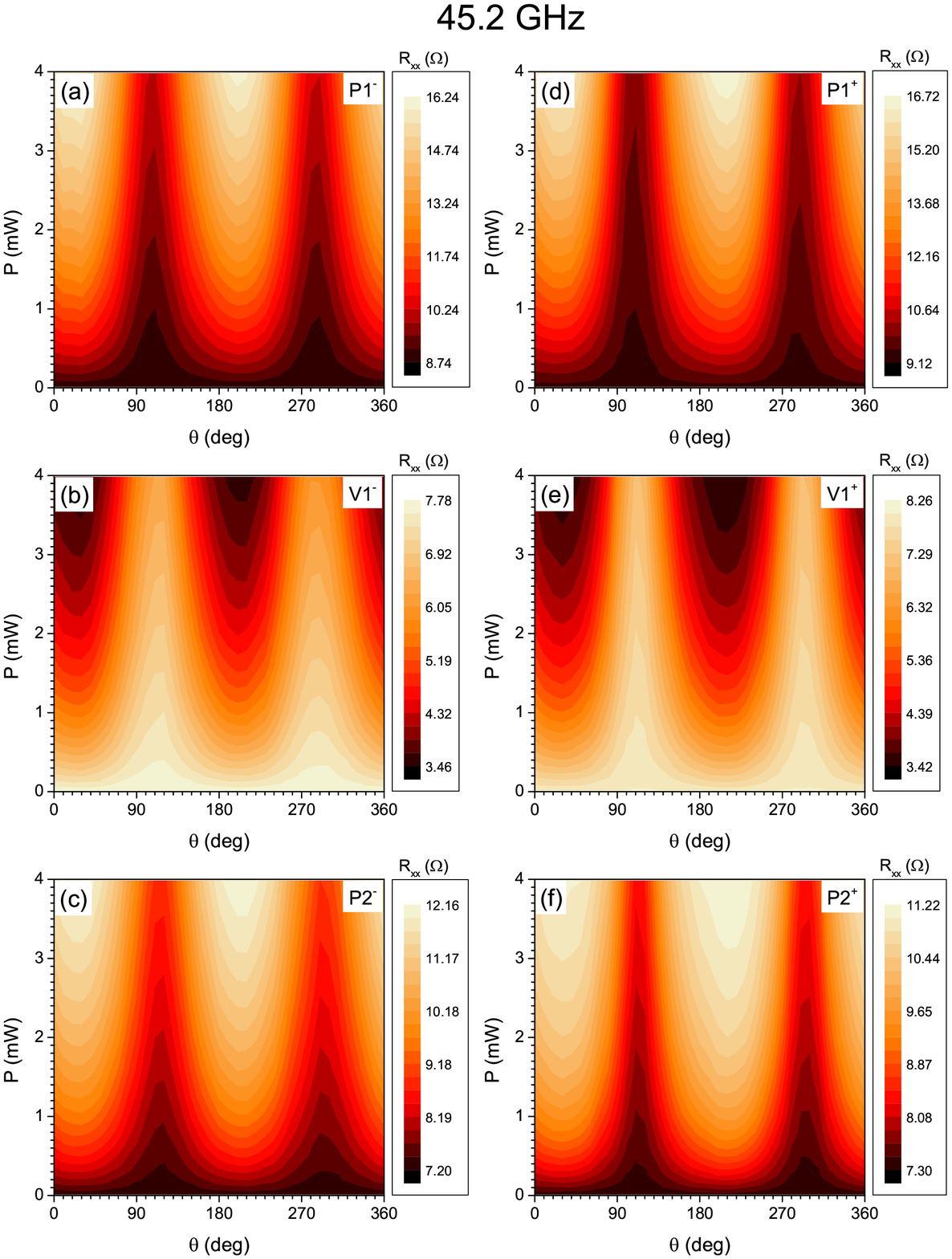}
\begin{center}
Figure 2
\end{center}
\end{figure}

\begin{figure}[t]
\centering
\includegraphics[width= 150mm]{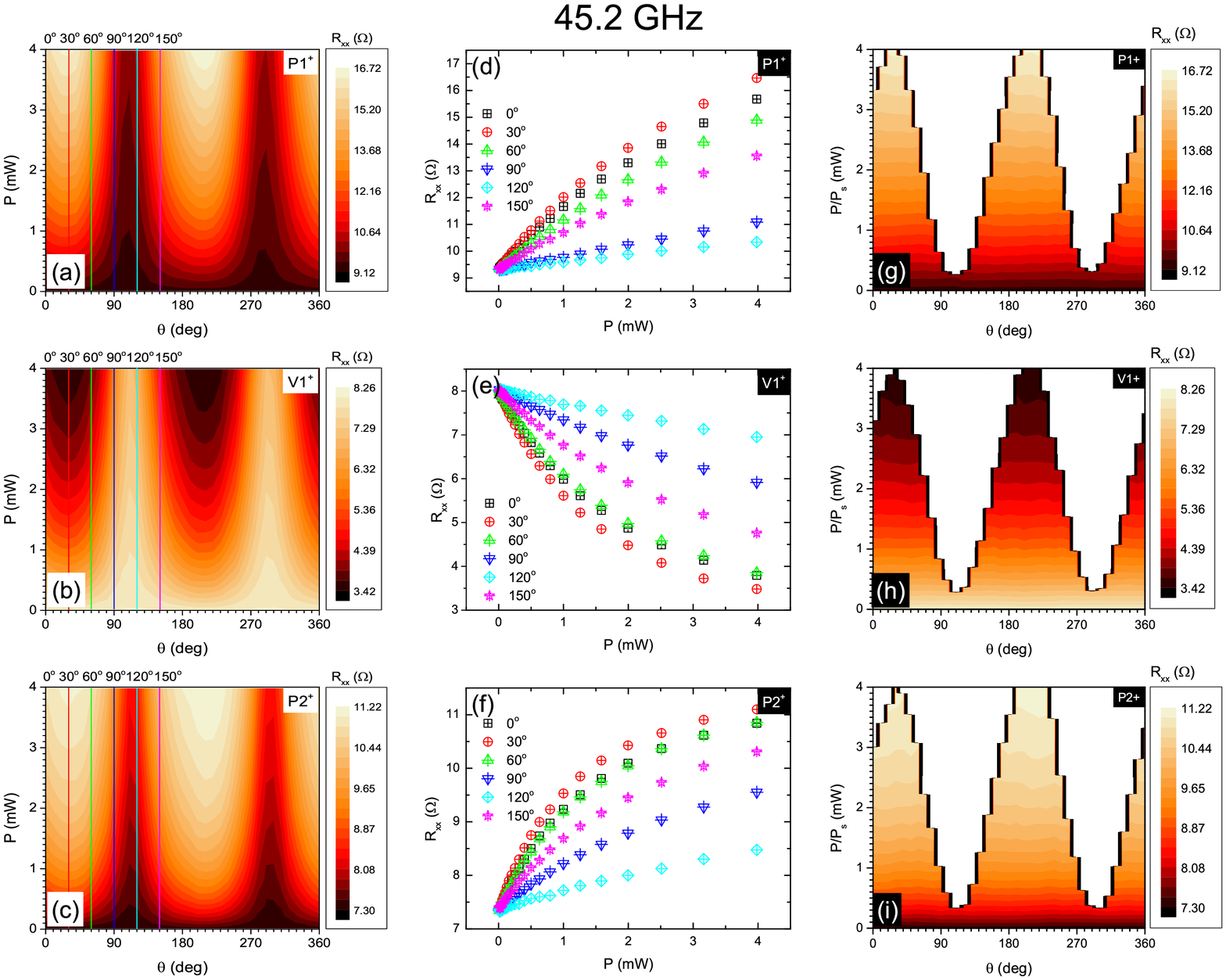}
\begin{center}
Figure 3
\end{center}
\end{figure}

\begin{figure}[t]
\centering
\includegraphics[width= 150mm]{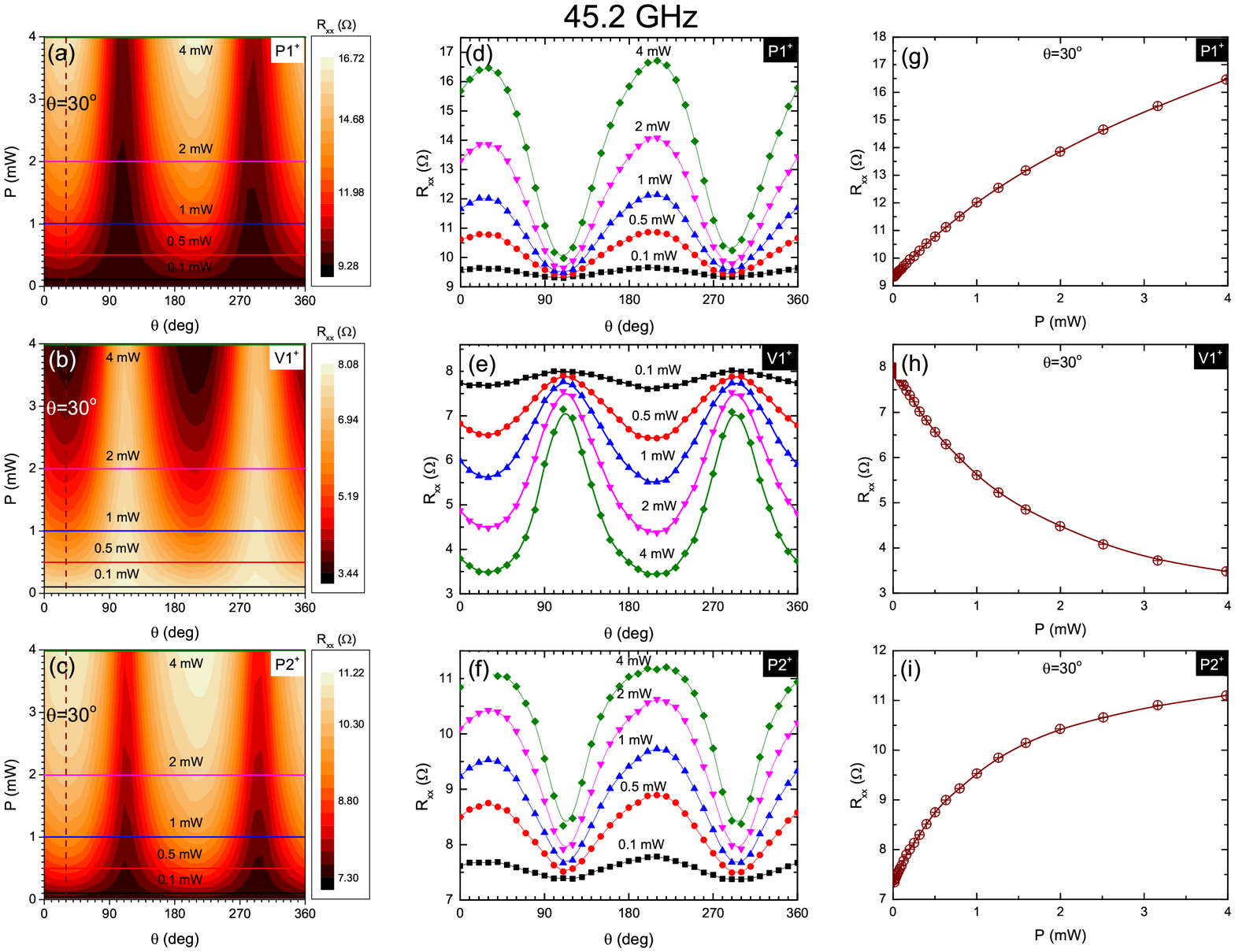}
\begin{center}
Figure 4
\end{center}
\end{figure}

\end{document}